# Axis-dependent carrier polarity in polycrystalline NaSn$_2$As$_2$

Naoto Nakamura, Yosuke Goto,[*] and Yoshikazu Mizuguchi

Department of Physics, Tokyo Metropolitan University, Hachioji 192-0397, Japan

[*]e-mail: y_goto@tmu.ac.jp

**Abstract**

Transverse thermoelectric devices consist of only one thermoelectric material, unlike conventional longitudinal thermoelectric devices that require two types of thermoelectric materials with p- and n-type polarities. However, scalable synthesis of materials that demonstrate axis-dependent carrier polarity, which is a prospective component to demonstrate the transverse thermoelectric device, is challenging. This paper reports that polycrystalline $NaSn_2As_2$, which was prepared by using uniaxial hot pressing, displayed axis-dependent carrier polarity. The preferred orientation of the sample was confirmed through X-ray diffraction measurements. Seebeck coefficient measurements indicate that carrier polarity depends on the measurement direction, which is consistent with recently reported results on single crystals of $NaSn_2As_2$. Given that our sample preparation procedure is readily scalable, the present work shows the possibility for preparing transverse thermoelectric devices using polycrystalline $NaSn_2As_2$ with a preferred orientation.

Thermoelectric devices are solid-state devices that convert heat flux directly into electrical energy without using any gas or liquid working fluid. They have been successfully used in space and other niche applications.[1-3] Conventional longitudinal thermoelectric devices require two types of thermoelectric materials with p- and n-type polarities that are connected via electrodes. Various research works have explored high-performance thermoelectric materials that demonstrate low electrical resistivity, low thermal conductivity, and a high Seebeck coefficient.[4,5] However, constructing longitudinal thermoelectric devices using these new materials is not simple, because apart from the above-mentioned transport properties, material properties also include chemical stability, mechanical strength, compatibility factor,[6] and so on.

Conversely, transverse thermoelectric device can be constructed by only one thermoelectric material, offering promise to expand the material choice and thermoelectric applications. The transverse thermoelectrics devices have been studied since the 1960s. However, research is still limited mainly due to insufficient progress in material development.[7-11] Transverse thermoelectric phenomena require breaking the directional symmetry of the Seebeck tensor. A promising approach is to use materials with axis-dependent carrier polarity, namely, p-type polarity in one direction and n-type orthogonal, resulting in off-diagonal terms that drive heat flow transverse to the electrical current.[12,13] Several materials that exhibit axis-dependent carrier polarity have been reported so far, such as Mg,[14] Zn,[14] Cd,[14] Re[15] and Be,[16] $CsBi_4Te_6$,[17] $Al_{13}Co_4$,[18] $Tl_2Ba_2CuO_{6-x}$,[19] $\kappa$-$(BEDT-TTF)_2Cu[N(CN)_2]Cl$,[20] $Re_4Si_7$,[21] $NaSnAs$,[22] and $NaSn_2As_2$.[23] Furthermore, first-principle calculation results predict several candidate materials for axis-dependent carrier polarity, although some of them have not yet been experimentally demonstrated.[24,25] However, despite recent theoretical and experimental progress in the field, scalable synthesis of materials that display axis-dependent carrier polarity has not yet been established.

Here, we demonstrate that polycrystalline $NaSn_2As_2$ prepared utilizing uniaxial hot pressing show axis-dependent carrier polarity. Axis-dependent carrier polarity of $NaSn_2As_2$ single crystals has been recently reported.[23] The axis-dependent polarity of $NaSn_2As_2$ is called goniopolarity as it stems from its unique Fermi surface geometry with a single band. The crystal structure of $NaSn_2As_2$ consists of a buckled honeycomb network of SnAs bound by the van der Waals forces and separated by Na ions, as schematically shown in Fig. 1.[26] $NaSn_2As_2$ and related materials have also been received great attention due to their intriguing properties as well as their goniopolarity. For example, nanometer-scale thickness sheets can be produced by exfoliation owing to the relatively weak van der Waals bonding between conducting layers.[27-29] NaSnAs demonstrated a low thermal conductivity, which is required for efficient thermoelectric materials. Emphasis was placed on the effects of lone-pair electrons.[30] Low thermal conductivity is also demonstrated in $Li_{1-x}Sn_{2+x}As_2$ and $Li_{1-x}Sn_{2+x}P_2$ due to Li/Sn mixed occupation with local ordering.[31,32] Superconducting transition at a low temperature was observed in $NaSn_2As_2$ and $Na_{1-x}Sn_2P_2$, meaning that these compounds can also be categorized as novel layered superconductors.[33-37] Because various elemental substitution is possible, these compounds are attractive to investigate further various functionalities. In this study, we demonstrate scalable synthesis and the axis-dependent carrier polarity of polycrystalline $NaSn_2As_2$.

We synthesized $NaSn_2As_2$ using Na (99.9%), Sn (99.99%), and As (99.9999%) as starting materials.[33] The raw elements were handled in an Ar-filled glovebox with a gas-purifier system. A surface oxide layer of Na was mechanically cleaved before experiments. The stoichiometric ratio of Na, Sn, and As was sealed in an evacuated quartz tube, heated at 700 °C–750 °C for 20 h, and cooled to 350 °C for 10 h, followed by furnace cooling. Uniaxial hot pressing was conducted using grinded $NaSn_2As_2$ and graphite die at 350 °C–450 °C at 70 MPa for 10 min (S. S. Alloy, PLASMAN CSP-KIT-02121). The size of grinded $NaSn_2As_2$ for hot pressing varied from several mm to below 100 μm (powder form). The hot-pressed samples, which were typically 7 mm in diameter and 8 mm in thickness, had a relative density greater than 90%.

We examined the phase purity of the sample using an X-ray diffractometer (XRD; Rigaku miniflex 600). Powder XRD was measured using grinded samples. XRD was also collected using the surface of hot-pressed pellet to evaluate the orientation degree. The orientation degree was evaluated using the Lotgering factor, $LF = \frac{(p-p_0)}{(1-p_0)}$, where $p$ denotes the summation fraction of the peak intensities corresponding to the preferred orientation axis to that of the summation of all diffraction peaks in oriented materials. $p_0$ is $p$ of a material with randomly orientation.[38,39] Here, $LF = 0$ and 1 correspond to a random and a perfect orientation, respectively.

The electrical resistivity $\rho$ and the Seebeck coefficient $S$ were measured using the four-probe method and the quasi-steady-state method in a He atmosphere (Advance Riko ZEM-3). The Hall coefficient $R_H$ was measured using the five-probe method with a physical property measurement system (PPMS; Quantum Design) under magnetic field lower than 3000 Oe. Here, five-probe method was employed to eliminate the parasitic resistance voltage due to asymmetry of sample shape, contact resistance, wire arrangement, etc. Note that low-field measurements were carried out to obtain $R_H$ because the Hall voltage versus magnetic field deviates from the linear relationship under a high magnetic field, as implied by previous studies on single crystals of $NaSn_2As_2$.[23]

Figure 2 shows XRD patterns of the representative sample (sample A in Table 1). Most diffraction peaks can be attributed to those of $NaSn_2As_2$, except for a weak peak due to elemental Sn, as denoted by the asterisk in Fig. 2. The intensity ratio significantly depends on diffraction measurements, namely, parallel or perpendicular to the uniaxial hot-pressing direction, as schematically shown in the inset of Fig. 2. In the parallel direction, the 110 diffraction is emphasized, while in the perpendicular direction, the 00$l$ diffraction is signified. These results confirm that the obtained samples exhibit preferred orientation, as expected from its layered structure. *LF* with these preferred axes was evaluated to be $LF_{110} = 0.11$ and $LF_{00l} = 0.28$.

Hereafter, measurement direction of transport properties is denoted as follows: parallel to the hot-pressing direction as cross-plane, while perpendicular to that one as in-plane. Figure 3 demonstrate the temperature dependence of the Seebeck coefficient in both directions. The cross-plane Seebeck coefficient has a positive value, indicating that the hole is a dominant carrier. On the other hand, the negative sign observed in the in-plane direction shows that the electron is a dominant carrier in this direction. These results are consistent with those observed in $NaSn_2As_2$ single crystals,[23] indicating that the axis-dependent carrier polarity is preserved in polycrystalline $NaSn_2As_2$. Notably, the observed in-plane Seebeck coefficient, $-8$ $\mu VK^{-1}$ at 300 K, agrees with measurement results on single crystals. However, the cross-plane Seebeck coefficient of 2 $\mu VK^{-1}$ is noticeably smaller than that of single crystals (8–10 $\mu VK^{-1}$). This is likely due to the fact that the orientation degree of the obtained samples is not perfect, as estimated using *LF*. We note that the observed Seebeck coefficient is almost independent from temperature in ranges between 300 and 370 K.

Figure 4 shows electrical resistivity as a function of temperature. In both measurement directions, very low electrical resistivities are detected, similarly to in-plane resistivity of single crystals.[23,33,35] We note that cross-plane resistivity is ~100 times higher than in-plane resistivity for single crystal.[23]

Table 1 summarizes the Seebeck coefficient of different batch samples. We examined several synthesis conditions, including the synthesis temperature (700 °C–750 °C), the hot-pressing temperature (350 °C–450 °C), and the size of raw $NaSn_2As_2$ for uniaxial hot pressing (several mm to powder form). The Seebeck coefficient slightly varies from sample to sample, however, the axis-dependent carrier polarity is reproducible in polycrystalline $NaSn_2As_2$.

We briefly comment on sample preparation temperature. In this study, $NaSn_2As_2$ was synthesized by the reaction of Na, Sn, and As at 700 °C–750 °C. After the synthesis at 750 °C, quartz tube was black colored, suggesting the raw materials also reacted with quartz tube. Therefore, we examined low temperature synthesis to avoid such an undesirable reaction. However, reaction temperature as high as 700 °C was required to obtain $NaSn_2As_2$ with sufficient quality. Lattice parameters slightly varied from sample to sample, but systematic trend with synthesis temperature was not observed. Notably, transport properties are likely independent of the reaction temperature, as summarized in Table 1.

Uniaxial hot pressing was conducted using grinded $NaSn_2As_2$ at 350 °C–450 °C. The sample pressed at 350 °C had a relative density of 90%. Because dense sample is required to characterize the intrinsic transport properties, the lower limit for hot-pressing temperature was determined to be 350 °C. The sample pressed at 500 °C did not show axis-dependent carrier polarity, likely because of insufficient preferred orientation. This seems to require the confirmation including reproducibility. It may be important to characterize the sample with and without axis-dependent carrier polarity in detail. Unfortunately, *LF*, which is used to evaluate the preferred orientation in this study, is not sufficient for this purpose because *LF* depends on measurement position even in the same batch sample, as shown in Figure S1.

The Hall coefficient of sample A is calculated to be $-3.5 \times 10^{-10}$ $m^3C^{-1}$ in the in-plane direction and $-1.7 \times 10^{-9}$ $m^3C^{-1}$ in the cross-plane direction. Although these values are close to those of single crystals, the signs are not consistent. In single crystals, the negative and positive signs were observed in the cross-plane and in-plane directions, meaning that the sign of the Hall coefficient is opposite to that of the Seebeck coefficient.[23] These results seem to be confirmed by first-principle calculations.[40] Again, the deviation from previous studies most probably arises from difference between single crystals and polycrystalline samples with preferred orientation.

In summary, we prepared polycrystalline $NaSn_2As_2$ with a preferred orientation using uniaxial hot pressing. Axis-dependent carrier polarity is confirmed by Seebeck coefficient measurements. As our sample preparation procedure is scalable, our work shows the possibility for preparing transverse thermoelectrics using polycrystalline $NaSn_2As_2$ with a preferred orientation, thereby expanding material choice for thermoelectric devices.

**Supplementary Material**

See supplementary material for XRD patterns of NaSn$_2$As$_2$ measured on different positions of the same batch sample.

**Acknowledgments**

This work was partly supported by JST CREST (No. JPMJCR16Q6), JSPS KAKENHI (No. 19K15291, 20KK0124), and Tokyo Metropolitan Government Advanced Research (H31-1).

**Data availability statement**

The data that support the findings of this study are available from the corresponding author upon reasonable request.

1. L. E. Bell, Science **321**, 1457 (2008).
2. A. F. Ioffe, *Semiconductor Thermoelements and Thermoelectric Cooling* (Infosearch, London, 1957).
3. C. B. Vining, Nat. Mater. **8**, 83 (2009).
4. G. J. Snyder and E. S. Toberer, Nat. Mater. **7**, 105 (2008).
5. J. R. Sootsman, D. Y. Chung, and M. G. Kanatzidis, Angew. Chem. Int. Ed. **48**, 8616 (2009).
6. G. J. Snyder, Appl. Phys. Lett. **84**, 2436 (2004).
7. A. A. Snarskii and L. P. Bulat, *Thermoelectrics Handbook, Macro to Nano* (CRC, Boca Raton, FL, 2006), Chap. 45.
8. Th. Zahner, R. Förg, and H. Lengfellner, Appl. Phys. Lett. **73**, 1364 (1998).
9. A. Kyarad and H. Lengfellner, Appl. Phys. Lett. **85**, 5613 (2004).
10. A. Kyarad and H. Lengfellner, Appl. Phys. Lett. **89**, 192103 (2006).
11. T. Kanno, S. Yotsuhashi, A. Sakai, K. Takahashi, and H. Adachi, Appl. Phys. Lett. **94**, 061917 (2009).
12. C. Zhou, S. Birner, Y. Tang, K. Heinselman, and M. Grayson, Phys. Rev. Lett. **110**, 227701 (2013).
13. Y. Tang, B. Cui, C. Zhou, and M. Grayson, J. Electron. Mater. **44**, 2095 (2015).
14. V. A. Rowe and P. A. Schroeder, Thermopower of Mg, Cd, and Zn between 1.2 degrees K and 300 degrees K. J. Phys. Chem. Solids **31**, 1 (1970).
15. A. T. Burkov, M. V. Vedernikov, V. A. Elenskii, G. P. Kovtun, Anisotropy of thermo-EMF and electroconductivity of high-purity rhenium. Fiz. Tverd. Tela **28**, 785 (1986).
16. A. T. Burkov, M. V. Verdernikov, Anomalous aniosotropy of high temperature thermo-EMF in beryllium. Sov. Phys. Solid State **28**, 3737 (1986).
17. D. Y. Chung, T. P. Hogan, M. Rocci-Lane, P. Brazis, J. R. Ireland, C. R. Kannewurf, M. Bastea, C. Uher, and M. G. Kanatzidis, J. Am. Chem. Soc. **126**, 6414 (2004).
18. J. Dolinšek, M. Komelj, P. Jeglič, S. Vrtnik, D. Stanić, P. Popčević, J. Ivkov, A. Smontara, Z. Jagličić, P. Gille, and Y. Grin, Phys. Rev. B **79**, 184201 (2009).
19. W. Kiehl, H. M. Duan, and A. M. Hermann, Physica C **253**, 271 (1995).
20. Y. Kawasugi, K. Seki, Y. Edagawa, Y. Sato, J. Pu, T. Takenobu, S. Yunoki, H. M. Yamamoto, R. Kato, Appl. Phys. Lett. **109**, 233301 (2016).
21. J. J. Gu, M. W. Oh, H. Inui, and D. Zhang, Phys. Rev. B. **71**, 113201 (2005).
22. A. M. Ochs, P. Gorai, Y. Wang, M. R. Scudder, K. Koster, C. E. Moore, V. Stevanovic, J. P. Heremans, W. Windl, E. S. Toberer, and J. E. Goldberger, Chem. Mater. (2021). DOI: 10.1021/acs.chemmater.0c04030

23. B. He, Y. Wang, M. Q. Arguilla, N. D. Cultrara, M. R. Scudder, J. E. Goldberger, W. Windl, and J. P. Heremans, Nat. Mater. **18**, 568 (2019).
24. K. P. Ong, D. J. Singh, and P. Wu, Phys. Rev. Lett. **104**, 176601 (2010).
25. Y. Wang, K. G. Koster, A. M. Ochs, M. R. Scudder, J. P. Heremans, W. Windl, and J. E. Goldberger, J. Am. Chem. Soc. **142**, 6 (2020).
26. M. Asbrand and B. Eisenmann, Z. Anorg. Allg. Chem. **621**, 576 (1995).
27. M. Q. Arguilla, J. Katoch, K. Krymowski, N. D. Cultrara, J. Xu, X. Xi, A. Hanks, S. Jiang, R. D. Ross, R. J. Koch, S. Ulstrup, A. Bostwick, C. Jozwiak, D. W. Mccomb, E. Rotenberg, J. Shan, W. Windl, R. K. Kawakami, and J. E. Goldberger, ACS Nano **10**, 9500 (2016).
28. M. Q. Arguilla, N. D. Cultrara, Z. J. Baum, S. Jiang, R. D. Ross, and J. E. Goldberger, Inorg. Chem. Front. **2**, 378 (2017).
29. G. M. Pugliese, F. Stramaglia, Y. Goto, K. Terashima, L. Simonelli, H. Fujiwara, A. Puri, C. Marini, M. Y. Hacisalihoglu, F. D'Acapito, T. Yokoya, T. Mizokawa, Y. Mizuguchi, and N. L. Saini, J. Phys. Condens. Matter **31**, 425402 (2019).
30. Z. Lin, G. Wang, C. Le, H. Zhao, N. Liu, J. Hu, L. Guo, and X. Chen, Phys. Rev. B **95**, 165201 (2017).
31. K. Lee, D. Kaseman, S. Sen, I. Hung, Z. Gan, B. Gerke, R. Po, M. Feygenson, J. Neuefeind, O. I. Lebedev, and K. Kovnir, J. Am. Chem. Soc. **137**, 3622 (2015).
32. Y. Goto, S. Nakanishi, Y. Nakai, T. Mito, A. Miura, C. Moriyoshi, Y. Kuroiwa, H. Usui, T. D. Matsuda, Y. Aoki, Y. Nakacho, Y. Yamada, K. Kanamura, Y. Mizuguchi, J. Mater. Chem. A **9**, 7034 (2021).
33. Y. Goto, A. Yamada, T. D. Matsuda, Y. Aoki, and Y. Mizuguchi, J. Phys. Soc. Jpn. **86**, 123701 (2017).
34. K. Ishihara, T. Takenaka, Y. Miao, O. Tanaka, Y. Mizukami, H. Usui, K. Kuroki, M. Konczykowski, Y. Goto, Y. Mizuguchi, and T. Shibauchi, Phys. Rev. B **98**, 20503 (2018).
35. E. J. Cheng, J. M. Ni, F. Q. Meng, T. P. Ying, B. L. Pan, Y. Y. Huang, Darren Peets, Q. H. Zhang, S. Y. Li, EPL **123**, 47004 (2018).
36. Y. Goto, A. Miura, C. Moriyoshi, Y. Kuroiwa, T. D. Matsuda, Y. Aoki, and Y. Mizuguchi, Sci. Rep. **8**, 12852 (2018).
37. H. Yuwen, Y. Goto, R. Jha, A. Miura, C. Moriyoshi, Y. Kuroiwa, D. Matsuda, Y. Aoki, and Y. Mizuguchi, Jpn. J. Appl. Phys. **58**, 083001 (2019).
38. F. K. Lotgering, J. Inorg. Nucl. Chem. **9**, 113 (1959).
39. R. Furushima, S. Tanaka, Z. Kato, and K. Uematsu, J. Ceram. Soc. Japan **118**, 921 (2010).

40. Y. Wang and P. Narang, Phys. Rev. B **102**, 125122 (2020).

Table 1

Cross-plane and in-plane Seebeck coefficient at 300 K for $NaSn_2As_2$ in different batches.

| Sample entry | Seebeck coefficient (μV/K) | |
|---|---|---|
| | Cross-plane | In-plane |
| A | 2 | −8 |
| B | 2 | −4 |
| C | 2 | −5 |
| D | 2 | −9 |
| E | 2 | −2 |
| F | 3 | −4 |

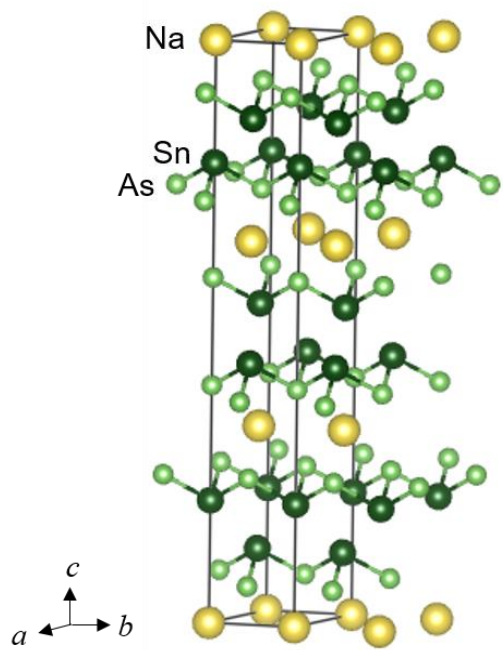

Figure 1
Crystal structure of NaSn$_2$As$_2$. Solid line represents the unit cell.

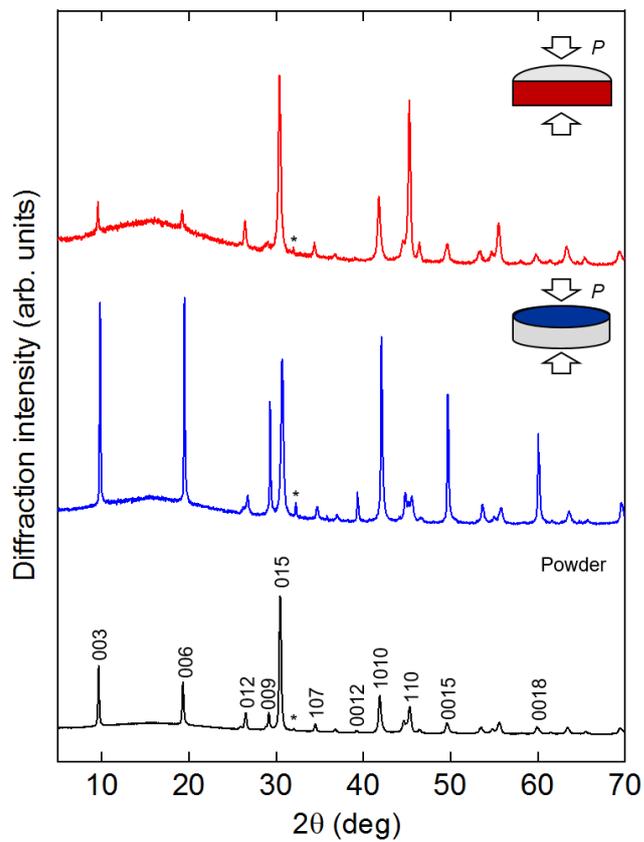

Figure 2

XRD patterns of NaSn$_2$As$_2$. Upper (middle) panel represent the measurement results on parallel (perpendicular) plane with respect to uniaxial hot pressing, as schematically shown in the inset. Powder XRD pattern of grinded sample is shown at the bottom. The asterisk denotes the diffraction peak due to elemental Sn.

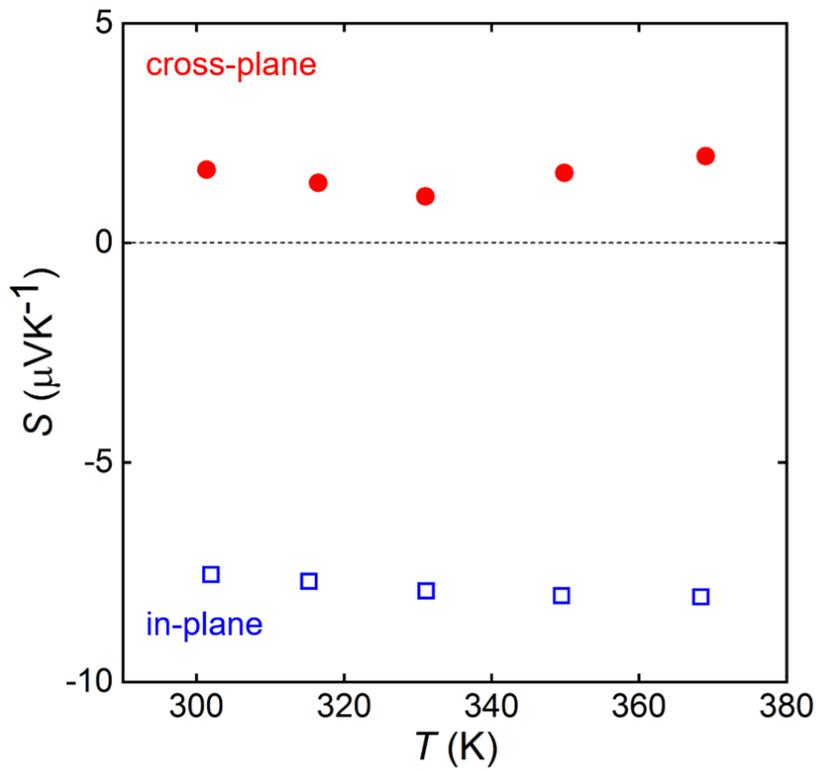

Figure 3

Cross-plane and in-plane Seebeck coefficient as a function of temperature for $NaSn_2As_2$.

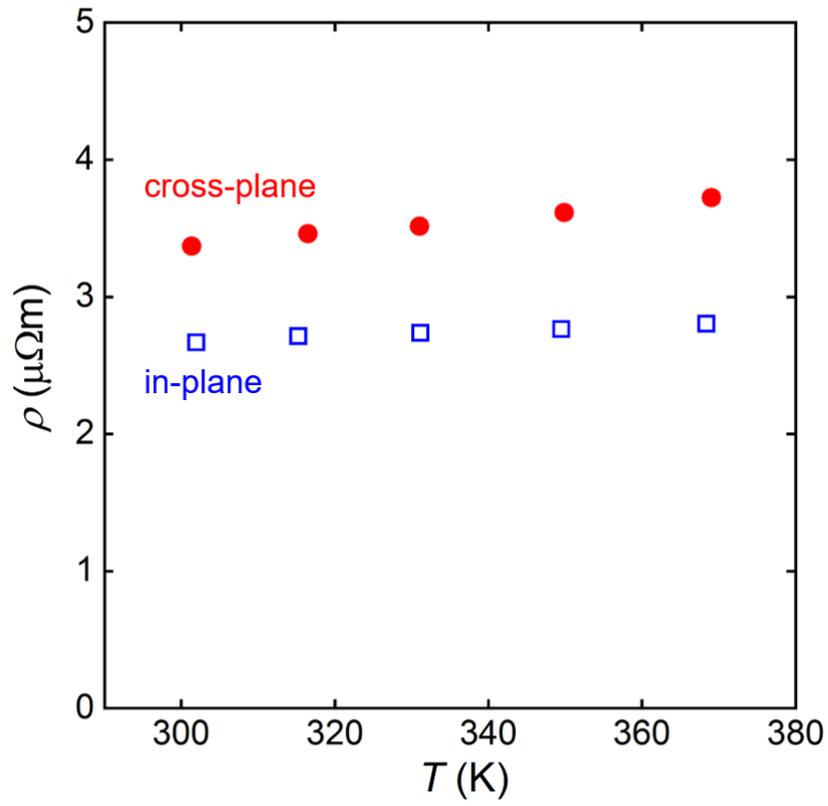

Figure 4

Cross-plane and in-plane electrical resistivity as a function of temperature for NaSn$_2$As$_2$.

Supporting Information for

# Axis-dependent carrier polarity in polycrystalline NaSn$_2$As$_2$

Naoto Nakamura, Yosuke Goto,* and Yoshikazu Mizuguchi

Department of Physics, Tokyo Metropolitan University, Hachioji 192-0397, Japan

*e-mail: y_goto@tmu.ac.jp

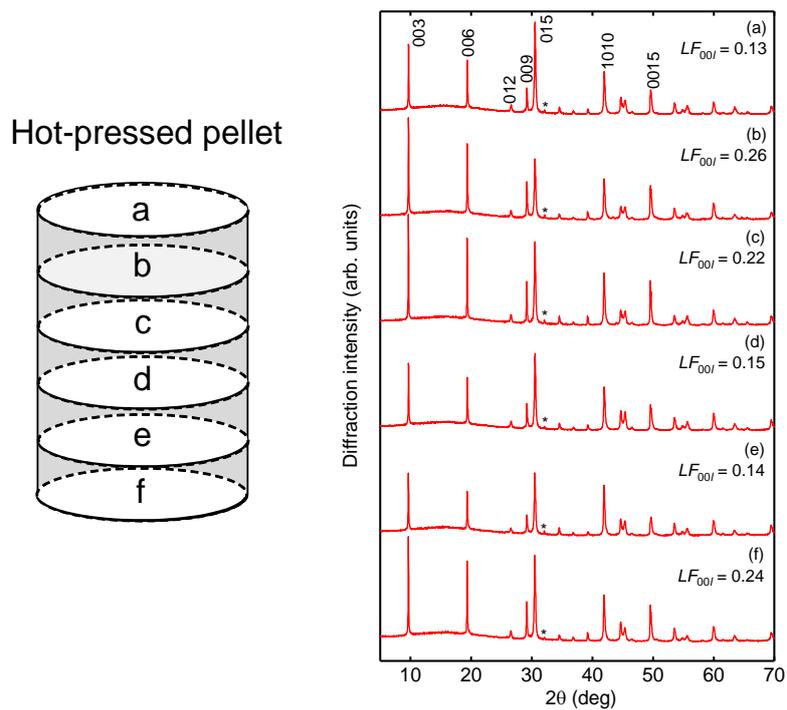

Figure S1

X-ray diffraction patterns of NaSn$_2$As$_2$ measured on perpendicular plane with respect to uniaxial hot pressing. Measurements are performed for different positions (a-f) of the same batch sample. Lotgering factor of (00$l$) planes ($LF_{00l}$) is also noted. The asterisk denotes the diffraction peak due to elemental Sn.